\documentclass[11pt,a4paper]{article}
\pdfoutput=1

\usepackage{amsmath}
\usepackage[T1]{fontenc}

\usepackage{jheppub}
\usepackage{psfrag}
\usepackage{slashed}
\usepackage{cancel}
\usepackage{lscape}
\usepackage{caption}
\usepackage{array}
\usepackage{graphicx}
\usepackage{subcaption}
\usepackage{multirow}
\usepackage{tabularx}
\usepackage{makecell}
\usepackage[utf8]{inputenc}
\usepackage{amsmath}
\usepackage{amssymb}
\usepackage{color}
\usepackage{slashed}
\usepackage{comment}
\usepackage{siunitx}[=v2]
\definecolor{mypink}{RGB}{219, 48, 122}
\definecolor{mygreen}{rgb}{0,0.7,0}
\definecolor{raspberry}{rgb}{0.53,0.15,0.34}

\def\la{\langle}
\def\ra{\rangle}
\def\spA#1#2{\la#1#2\ra}
\def\spB#1#2{[#1#2]}

\DeclareMathOperator{\tr}{\rm tr}

\def\eps{\epsilon}

\newcommand{\njet}{\texttt{NJet}}
\newcommand{\pentagonfunctions}{\texttt{PentagonFunctions++}}
\newcommand{\finiteflow}{\texttt{FiniteFlow}}
\newcommand{\qd}{\texttt{QD}}
\newcommand{\eigen}{\texttt{Eigen}}
\newcommand{\nnlojet}{\texttt{NNLOJET}}
\newcommand{\mathematica}{\texttt{Mathematica}}
\newcommand{\form}{\texttt{FORM}}
\newcommand{\qgraf}{\texttt{QGRAF}}
\newcommand{\spinney}{\texttt{Spinney}}
\newcommand{\cpp}{\texttt{C++}}

\newcommand{\fp}{\texttt{f64}}
\newcommand{\fpp}{\texttt{f128}}
\newcommand{\fppp}{\texttt{f256}}

\def\hpl11{{\mathrm{HPL}}_{1,1}}

\newcolumntype{C}[1]{>{\hsize=#1\hsize\centering\arraybackslash}X}%

\newcolumntype{Z}{r<{\hspace{3mm}}}
\title{Virtual QCD corrections to gluon-initiated diphoton plus jet production at hadron colliders}

\author[a]{Simon Badger,}
\author[b]{Christian Br\o nnum-Hansen,}
\author[c]{Dmitry Chicherin,}
\author[d]{Thomas Gehrmann,}
\author[e]{Heribertus Bayu Hartanto,}
\author[f]{Johannes Henn,}
\author[d]{Matteo Marcoli,}
\author[g]{Ryan Moodie,}
\author[h]{Tiziano Peraro,}
\author[a]{Simone Zoia}

\affiliation[a]{
Dipartimento di Fisica and Arnold-Regge Center, Universit\`{a} di Torino,
and INFN, Sezione di Torino, Via P. Giuria 1, I-10125 Torino, Italy
}
\affiliation[b]{
Institute for Theoretical Particle Physics, KIT, Karlsruhe, Germany
}
\affiliation[c]{
LAPTh, CNRS – USMB, BP 110, F-74941 Annecy-le-Vieux, France
}
\affiliation[d]{
Physik-Institut, Universit\"{a}t Z\"{u}rich, Wintherturerstrasse 190, CH-8057 Z\"{u}rich, Switzerland
}
\affiliation[e]{
Cavendish Laboratory, University of Cambridge, Cambridge CB3 0HE, United Kingdom
}
\affiliation[f]{
Max-Planck-Institut f\"{u}r Physik, Werner-Heisenberg-Institut, D-80805 M\"{u}nchen, Germany
}
\affiliation[g]{
Institute for Particle Physics Phenomenology, Department of Physics, Durham University, Durham DH1 3LE, United Kingdom
}
\affiliation[h]{
Dipartimento di Fisica e Astronomia, Universit\`a di Bologna e INFN, Sezione di Bologna, via Irnerio 46, I-40126 Bologna, Italy
}
\emailAdd{
simondavid.badger@unito.it
}

\abstract{
We present an analytic computation of the gluon-initiated contribution to diphoton plus jet production at hadron colliders up to two loops in QCD. We
reconstruct the analytic form of the finite remainders from numerical evaluations over finite fields including all colour contributions.  Compact
expressions are found using the pentagon function basis.  We provide a fast and stable implementation for the colour- and helicity-summed interference
between the one-loop and two-loop finite remainders in \cpp~as part of the \njet~library.
}

\preprint{CAVENDISH-HEP-21/10, IPPP/20/115, \\ \vspace{-0.3cm} \hspace{8cm} MPP-2021-86, TTP21-019, ZU-TH 26/21}

\begin{document}
\maketitle
\flushbottom

\section{Introduction}

Precise theoretical predictions are in high demand for the current Large Hadron Collider (LHC) experiments which are aiming to look for tiny
deviations from the Standard
Model (SM). Due to the relatively large size of the strong coupling constant, next-to-next-to-leading order (NNLO) corrections in quantum chromo-dynamics
(QCD) are desirable for a wide variety of final state processes. In particular, a class of $2\to 3$ scattering processes with many kinematic scales have
presented a considerable challenge to the theoretical community and there has been a good deal of activity leading to new methods able of overcoming their algebraic and analytic complexity~\cite{Kosower:2011ty,Mastrolia:2011pr,Badger:2012dp,Zhang:2012ce,Mastrolia:2012an,Mastrolia:2012wf,Ita:2015tya,Badger:2013gxa,Badger:2015lda,Abreu:2017xsl,Abreu:2020xvt}.

The production of a pair of high energy photons is an important experimental signature at hadron colliders and can be used for example to study the Higgs boson through its decay to photons. The SM backgrounds are dominated by QCD corrections and a precise description of the kinematics of these observables
requires the theoretical predictions to include perturbative information from the production in association with additional jets. NNLO corrections to
the process $pp\to \gamma\gamma+j$, which is initiated at LO by quark-antiquark and quark-gluon processes, have been considered a high priority for current and future experiments for several years~\cite{Badger:2016bpw,Bendavid:2018nar,Amoroso:2020lgh}, and were computed most recently~\cite{Chawdhry:2021hkp}. The Born-level amplitude for gluon-initiated diphoton plus jet production contains a
closed quark loop coupling to both photons. Consequently, this type of process starts to contribute to the cross section only from NNLO onwards. Owing to the large gluon luminosity, it yields a dominant contribution to the NNLO corrections and dominates their scale uncertainty~\cite{Chawdhry:2021hkp}. 
To improve upon this uncertainty requires the NLO corrections to the closed quark-loop contributions, which amount to the two-loop virtual amplitudes for
$gg \to \gamma\gamma g$ that we derive in this article. Curiously, the gluon channel has the opposite structure to the conventional expansion in the
number of colour charges, $N_c$. The dominant, leading colour, contributions to the quark-initiated process contain only planar diagrams, while in the gluon-initiated case the leading-colour limit contains both planar and non-planar graphs at two loops. Graphs with the highest complexity are thus unavoidable.

The last few years have seen rapid progress in our ability to compute two-loop $2\to3$ scattering processes in QCD which had been intractable for a
long time. The analytic computation of the scattering amplitudes in a form suitable for phenomenological applications requires a number of major technical
bottlenecks to be overcome. A basis of special functions must be identified that can be evaluated efficiently over the full phase space. For massless
five-point scattering, such a basis has been identified~\cite{Chicherin:2017dob,Papadopoulos:2015jft,Gehrmann:2018yef,Chicherin:2018mue,Chicherin:2018old} and became recently available as a fast and stable
implementation in \texttt{C++} valid in the physical scattering region~\cite{Chicherin:2020oor}. Secondly, the amplitude must be reduced from tensor
Feynman integrals onto a basis of master integrals that can subsequently be expanded in terms of special functions. Currently, the only viable approach to
this task is through the solution of enormous systems of integration-by-parts (IBP) identities~\cite{Tkachov:1981wb,Chetyrkin:1981qh,Laporta:2001dd}
of which many public implementations now exist~\cite{Anastasiou:2004vj,Studerus:2009ye,vonManteuffel:2012np,Lee:2012cn,Smirnov:2019qkx,Klappert:2020nbg}.
There has been success in simplifying this problem using syzygy relations~\cite{Gluza:2010ws,Schabinger:2011dz,Ita:2015tya,Larsen:2015ped,Boehm:2017wjc},
module intersection~\cite{Boehm:2018fpv,Boehm:2020ijp},
intersection theory~\cite{Mastrolia:2018uzb,Frellesvig:2019uqt,Frellesvig:2019kgj,Frellesvig:2020qot},
$\eta$ expansion~\cite{Liu:2017jxz,Liu:2018dmc,Guan:2019bcx,Zhang:2018mlo,Wang:2019mnn},
direct solution of IBPs through recursive relations~\cite{Kosower:2018obg},
multivariate partial fractioning~\cite{Boehm:2020ijp}, and by-passing complicated algebraic steps through finite field arithmetic~\cite{vonManteuffel:2014ixa,Peraro:2016wsq,Klappert:2019emp,Peraro:2019svx,Klappert:2020aqs}.
The latter method can be applied more broadly~\cite{Peraro:2016wsq,Peraro:2019svx}, in particular to a complete reduction of the amplitudes into a representation using special functions.
In this article, we approach the problem through a direct analytic reconstruction of the amplitudes at the level of the pentagon
functions performing all intermediate steps numerically over finite fields. This technique has been applied successfully to leading-colour (planar) five-parton amplitudes first numerically~\cite{Badger:2017jhb,Abreu:2017hqn,Badger:2018gip,Abreu:2018jgq} and then
analytically~\cite{Gehrmann:2015bfy,Badger:2018enw,Abreu:2018zmy,Abreu:2019odu,Abreu:2021fuk}. Leading-colour three-photon production has also been completed and cross checked by
two independent groups both at the level of the amplitudes~\cite{Abreu:2020cwb,Chawdhry:2020for} and of differential cross
sections~\cite{Chawdhry:2019bji,Kallweit:2020gcp}. 
Very recently, NNLO QCD predictions for a number of three-jet observables and differential three-to-two jet ratios have been computed at leading colour as well~\cite{Czakon:2021mjy}.  
The process $gg\to g\gamma\gamma$ contains the most complicated non-planar topologies with up to rank five tensor numerators even at leading colour.

Diphoton production has been known at NNLO for some time~\cite{Catani:2011qz,Campbell:2016yrh} and the two-loop scattering amplitudes were among the first complete $2\to 2$ process to be
calculated~\cite{Anastasiou:2002zn,Bern:2001df}. The first results for the amplitudes for $pp\to \gamma\gamma j$ appeared in the last few months
both for the amplitudes~\cite{Agarwal:2021grm,Chawdhry:2021mkw} and NNLO differential cross section~\cite{Chawdhry:2021hkp} in the leading-colour
approximation. Very recently, the full-colour two-loop QCD corrections for the quark-initiated channels to $pp\to\gamma\gamma j$ were presented~\cite{Agarwal:2021vdh}.

We obtain sufficiently compact analytic expressions for the complete set of helicity amplitudes for which the ultraviolet (UV) and
infrared (IR) poles have been subtracted, and implement them into an efficient and stable \texttt{C++} code as part of the \njet~library~\cite{Badger:2012pg}.
These expressions take the form of rational coefficients multiplied by pentagon functions. The code provides colour- and helicity-summed expressions for the two-loop amplitudes interfered with the one-loop amplitudes, which can be used directly in phenomenological applications.

Our paper is organised as follows. We first introduce the notation and describe the colour decomposition of the amplitudes. We then describe the methodology used to perform the integration-by-parts reduction and reconstruction of the finite remainders over finite fields. In particular we describe a method for performing a univariate partial fractioning of the rational coefficients of the special functions on the fly. This approach can be used inside the finite field workflow, reducing significantly the number of sample points required to complete the analytic reconstruction and yielding compact analytic expressions. In particular, we show explicitly some remarkably simple analytic forms we obtained for the all-plus helicity amplitude, which highlight its conformal properties.
Finally, we present the implementation in the \njet~library~\cite{Badger:2012pg} and the performance of the code using a realistic set of phase-space points before concluding with a few remarks on future applications of the results and methods. We also include an appendix with some details of the momentum twistor formalism used to provide a rational parametrisation of the kinematics.

\section{Kinematics and amplitude conventions}
\label{sec:Kinematics}

We consider the production of a pair of photons in association with a gluon from gluon fusion,
\begin{align}
g(-p_1) + g(-p_2) \to g(p_3) + \gamma(p_4) + \gamma(p_5) \, ,
\end{align}
up to two-loop order in QCD. All particles are massless, $p_i^2=0$, and we take all momenta as outgoing, so that
\begin{align}
\sum_{i=1}^5 p_i = 0\,.
\end{align}
Without loss of generality, we assume that the external momenta $p_i$ live in a four-dimensional Minkowski space-time, whereas the Feynman loop integrations are done in $d=4-2 \eps$ to regulate the divergences.
The kinematics are described by five independent scalar invariants, which can be chosen as $\{s_{12}, s_{23}, s_{34}, s_{45}, s_{15} \}$ with $s_{ij} = (p_i + p_j)^2$, and a pseudo-scalar invariant,
\begin{align}
\text{tr}_5 = 4 i \epsilon_{\mu \nu \rho \sigma} p_1^{\mu} p_2^{\mu} p_3^{\mu} p_4^{\mu} = [12] \spA 23 [34] \spA 41 - \spA 12 [23] \spA 34 [41] \, .
\end{align}
The square of $\text{tr}_5$ can be expressed in terms of the scalar invariants through the Gram determinant of the external momenta,
\begin{align}
\text{tr}_5^2 = \Delta :=  \text{det}\left(2 p_i \cdot p_j \right)_{i,j=1,\ldots,4} \,,
\end{align}
which is a degree-4 polynomial in the $s_{ij}$. The pseudo-scalar invariant $\text{tr}_5$ therefore introduces an algebraic dependence on the kinematics, since $\text{tr}_5 = \pm \sqrt{\Delta}$. We emphasise that the sign of $\text{tr}_5$ changes under parity conjugation, which acts by flipping the sign of the spatial momentum components,
\begin{align}
P : \left(p_i^0 , \vec{p}_i\right) \longrightarrow  \left(p_i^0 , -\vec{p}_i\right) \, ,
\end{align}
and under odd-signature permutations of the external momenta.

We work in the $s_{12}$ physical scattering region, which is delimited by the requirements that all $s$-channel invariants are positive and all $t$-channel invariants are negative, 
\begin{align}
& s_{12}, s_{34}, s_{35}, s_{45} > 0 \,, \\
& s_{13}, s_{14}, s_{15}, s_{23}, s_{24}, s_{25} < 0 \,,
\end{align}
together with the negativity of the Gram determinant, $\Delta<0$, which follows from the real-valuedness of the momenta~\cite{Gehrmann:2018yef}.

The scattering of gluons and photons is a one-loop process at leading order. We decompose the scattering amplitude as
\begin{align}
  \mathcal{A}(1_g,2_g,3_g,4_\gamma,5_\gamma) = g_s g_e^2 \left( Q_u^2 N_u + Q_d^2 N_d \right) f^{a_1 a_2 a_3}
  \sum_{\ell=1}^{\infty} \left(n_\eps \frac{\alpha_s}{4\pi} \right)^{\ell} A^{(\ell)}(1_g,2_g,3_g,4_\gamma,5_\gamma) \,,
  \label{eq:colourdecomp}
\end{align}
where $n_{\eps} = i(4\pi/\mu_R^2)^\eps e^{-\eps \gamma_E}$ with $\mu_R$ being the renormalisation scale. In Eq.~\eqref{eq:colourdecomp}, $g_s$ and
$g_e$ are the strong and electromagnetic coupling constants, $\alpha_s = g_s^2/(4\pi)$, $N_q$ and $Q_q$ are the number of quarks of type $q$ and their
electric charge in units of the electron charge, and $a_i$ is the adjoint $\mathrm{SU}(N_c)$ colour index of the $i^\text{th}$ gluon.
The one-loop amplitude can be obtained from permutations of pure gluon scattering~\cite{Dicus:1987fk,deFlorian:1999tp}.

\begin{figure*}[t!]
    \centering
    \includegraphics[width=\textwidth]{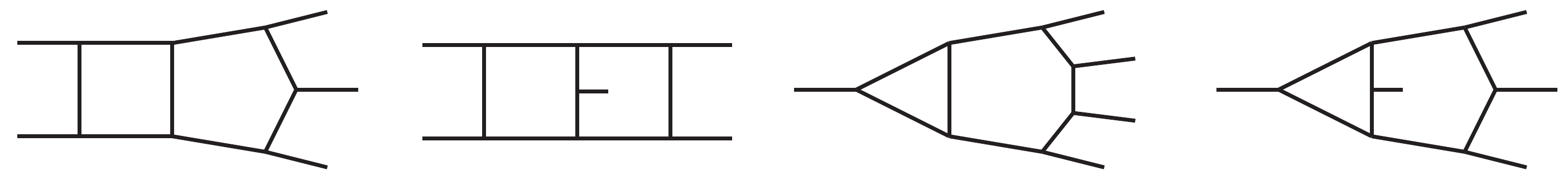}
    \caption{Independent integral families for the $gg \to g \gamma \gamma$ amplitude. The non-planar topologies appear only in the leading-colour amplitude.}
    \label{fig:integral_topologies}
\end{figure*}

\begin{figure*}[t!]
    \centering
    \begin{subfigure}[t]{0.18\textwidth}
        \centering
        \includegraphics[height=1.2cm]{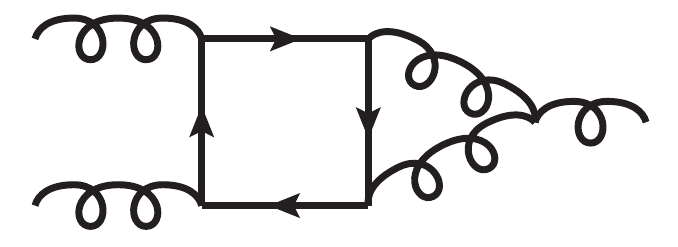}
        \caption{$N_c$}
    \end{subfigure}
    ~
    \begin{subfigure}[t]{0.18\textwidth}
        \centering
        \includegraphics[height=1.2cm]{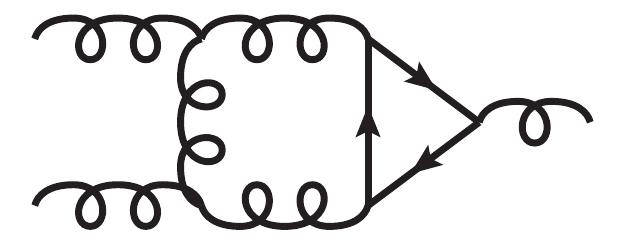}
        \caption{$N_c$}
    \end{subfigure}
    ~
    \begin{subfigure}[t]{0.18\textwidth}
        \centering
        \includegraphics[height=1.3cm]{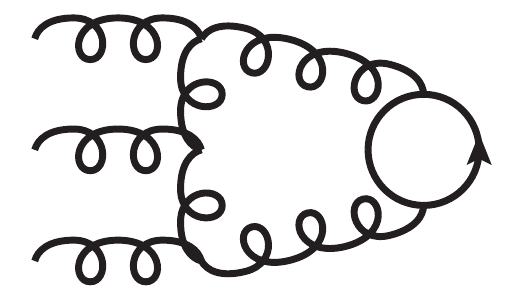}
        \caption{$N_c$}
    \end{subfigure}
    ~
    \begin{subfigure}[t]{0.18\textwidth}
        \centering
        \includegraphics[height=1.2cm]{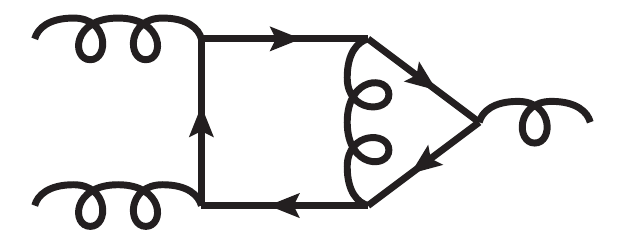}
        \caption{$\frac{1}{N_c}$}
    \end{subfigure}
    ~
    \begin{subfigure}[t]{0.18\textwidth}
        \centering
        \includegraphics[height=1.3cm]{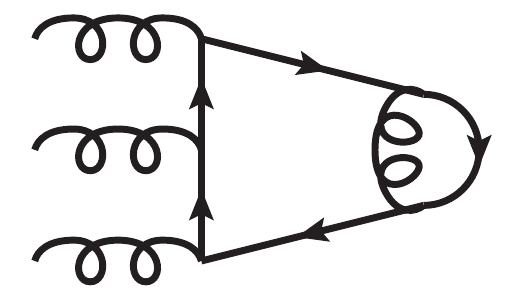}
        \caption{$N_c - \frac{1}{N_c}$}
    \end{subfigure}
    \caption{The colour factor of each diagram in the $gg \to g \gamma \gamma$ follows from the representative three-gluon, two-loop diagrams with a closed fermion loop shown here.}
    \label{fig:diagram_colours}
\end{figure*}

We further expand the loop amplitudes in powers of $N_c$ and $n_f$ (the number of light flavour fermions),
\begin{equation} \label{eq:Nc_exp}
\begin{aligned}
  A^{(1)}(1_g,2_g,3_g,4_\gamma,5_\gamma) &= A^{(1)}_1(1_g,2_g,3_g,4_\gamma,5_\gamma) \,, \\
  A^{(2)}(1_g,2_g,3_g,4_\gamma,5_\gamma) &= N_c A^{(2)}_1(1_g,2_g,3_g,4_\gamma,5_\gamma) \\
  &+ \frac{1}{N_c} A^{(2)}_2(1_g,2_g,3_g,4_\gamma,5_\gamma)
  + n_f A^{(2)}_3(1_g,2_g,3_g,4_\gamma,5_\gamma)
  \,.
\end{aligned}
\end{equation}
Surprisingly, the subleading-colour two-loop amplitudes contain only planar integrals, while the leading colour contains all of the four
independent families shown in Figure~\ref{fig:integral_topologies}. This pattern is the opposite to that of the quark-initiated channels computed in Refs.~\cite{Agarwal:2021grm,Chawdhry:2021mkw,Agarwal:2021vdh}, for which the leading-colour contributions involve only the planar integrals and are therefore simpler to compute. Providing a prediction for the gluon-initiated channel necessarily requires handling the most complicated integral families.
A simple analysis of the colour factors of each of the three-gluon vertex diagrams shown in Figure~\ref{fig:diagram_colours} illustrates how this pattern arises. Photons couple to any of the fermion propagators, and the colour factors remain the same. It can then be seen that non-planar contributions can come from the diagrams (a)--(c) only. Diagrams (d)--(e), which contribute to the subleading colour, remain planar (allowing for permutations of the external momenta).

In our setup, we reduce directly to the finite remainder where the UV and IR poles have been subtracted analytically. The poles take a particularly
simple form since there is no tree-level process and the one-loop amplitudes are finite in $\eps$. The one- and two-loop finite remainders are given
in terms of the bare amplitudes by~\cite{Catani:1998bh,Becher:2009qa,Becher:2009cu,Gardi:2009qi,Gardi:2009zv},
\begin{equation} \label{eq:finremdef}
\begin{aligned}
  F^{(1)} &= A^{(1)}(1_g,2_g,3_g,4_\gamma,5_\gamma) \,, \\
  F^{(2)} &= A^{(2)}(1_g,2_g,3_g,4_\gamma,5_\gamma) - \left(I^{(1)} + \frac{3}{2} \frac{\beta_0}{\eps} \right)A^{(1)}(1_g,2_g,3_g,4_\gamma,5_\gamma) \,,
\end{aligned}
\end{equation}
where $\beta_0 = 11 N_c/3 - 2 n_f/3$ and
\begin{align}
  I^{(1)} = - n_\Gamma(\eps) \, \left\{ \frac{N_c}{\eps^2} \left[ \left(\frac{\mu_R^2}{-s_{12}} \right)^\eps+\left(\frac{\mu_R^2}{-s_{23}}
  \right)^\eps+\left(\frac{\mu_R^2}{-s_{31}} \right)^\eps \right]  + 3 \frac{\gamma_g}{\eps} \right\} \,,
  \label{eq:I1}
\end{align}
with $n_\Gamma(\eps)=e^{\eps \gamma_{E}}/\Gamma(1-\eps)$ and $\gamma_g = \beta_0/2$ in the 't Hooft-Veltman scheme. The logarithms arising from the $\eps$-expansion of $I^{(1)}$ can be analytically continued to the $s_{12}$ channel by adding a small positive imaginary part to each $s_{ij}$. The $\beta_0$ term in the definition of the two-loop finite remainder accounts for the strong coupling renormalisation.
The finite remainders inherit from the amplitudes the decomposition in powers of $N_c$ and $n_f$ given by Eq.~\eqref{eq:Nc_exp},
\begin{equation} \label{eq:F_Nc_exp}
\begin{aligned}
  F^{(1)}(1_g,2_g,3_g,4_\gamma,5_\gamma) &= F^{(1)}_1(1_g,2_g,3_g,4_\gamma,5_\gamma) \,, \\
  F^{(2)}(1_g,2_g,3_g,4_\gamma,5_\gamma) &= N_c F^{(2)}_1(1_g,2_g,3_g,4_\gamma,5_\gamma) \\
  &+ \frac{1}{N_c} F^{(2)}_2(1_g,2_g,3_g,4_\gamma,5_\gamma)
  + n_f F^{(2)}_3(1_g,2_g,3_g,4_\gamma,5_\gamma)
  \,.
\end{aligned}
\end{equation}

Our final results are presented in the 't Hooft-Veltman scheme, although we make the distinction between the dimension $d$ of the loop integration and the dimension $d_s=g^\mu_{\ \mu}$ arising from the numerator algebra. Amplitudes with $d_s=2$ have a much simpler algebraic structure and contain information that can then be used to reduce the complexity of the more difficult $d_s-2$ component (see e.g.~Section~\ref{sec:LinearRelations}). The one-loop finite remainder has only the $d_s=2$ term, and we expand the two-loop finite remainder around $d_s=2$ as
\begin{align} \label{eq:ds_exp}
  F^{(2)}_k &= F^{(2)}_{k;0} + F^{(2)}_{k;1} \, (d_s-2)  \,.
\end{align}
The 't Hooft-Veltman scheme is obtained by setting $d_s = d = 4-2 \eps$.

\section{Computational setup and amplitude reduction}
\label{sec:Reduction}

We take a diagrammatic approach to the calculation of the amplitude along the lines of previous work~\cite{Hartanto:2019uvl,Badger:2021owl}.  Here we
briefly summarise the steps and refer the reader to Ref.~\cite{Badger:2021owl} for details.  All Feynman diagrams are generated using
\qgraf~\cite{Nogueira:1991ex} and subsequently processed using a combination of in-house \mathematica~and
\form~\cite{Kuipers:2012rf,Ruijl:2017dtg} scripts. In total, including contributions from ghost diagrams, we find 50 diagrams at one loop and
1527 at two loops. Aided by the \spinney~\cite{Cullen:2010jv} package to perform the 't Hooft algebra, the numerators are written for each
independent helicity configuration. From the loop denominator structure we assign an integral topology to each diagram.  At this point, the diagram
numerators are linear combinations of monomials in loop-momenta dependent scalar and spinor products with coefficients depending only on external
momenta.  These coefficients are loaded into a dataflow graph using \finiteflow~\cite{Peraro:2019svx}.  This enables numerical sampling over
finite fields, thus sidestepping analytically complicated intermediate expressions in further steps.  We rewrite loop-momenta dependent monomials into
inverse propagator denominators and a choice of irreducible scalar products (ISPs).  The required mapping of the coefficients is performed numerically
within the dataflow framework.  After summing all diagrams and dropping scaleless integrals, we arrive at an expression ready for integration-by-parts
(IBP) reduction.

The reduction to master integrals has been obtained using an improved version of the Laporta algorithm~\cite{Laporta:2001dd}.  For most integral
families we generated identities containing no higher power of propagators with respect to those appearing in the amplitude, following ideas proposed
in~\cite{Gluza:2010ws,Ita:2015tya,Larsen:2015ped}. These identities have been found using the Baikov representation of loop integrals, for which
identities (i) without higher powers of propagators and (ii) without dimension-shifted integrals can be found by solving polynomial equations called
syzygy equations. Closed form solutions to both these constraints are separately known.  Indeed, the solution of (i) is almost trivial and the solution for
(ii) has been found in Ref.~\cite{Boehm:2017wjc}.  The two syzygy solutions need to be combined for generating identities that satisfy both
constraints.  For this purpose we used a custom syzygy solver that implements the algorithm in Ref.~\cite{Schabinger:2011dz} using
\finiteflow~\cite{Peraro:2019svx}.  More details on this method can be found in
Refs.~\cite{Gluza:2010ws,Ita:2015tya,Larsen:2015ped,Boehm:2017wjc}.

For each integral family, we generated integral identities only for one permutation of the external legs.  Numerical solutions for all the
permutations contributing to an amplitude have been found by solving the systems of equations several times, with different numerical inputs for the
invariants.  Mappings between master integrals with different permutations of external legs are applied afterwards to obtain a result in terms of a
minimal set of them.

As an additional improvement, for each phase-space point evaluated on a finite field, we reconstruct the full dependence on the dimensional regulator
$\epsilon$ of the amplitude reduced to master integrals before substituting their expressions in terms of special functions and computing the Laurent
expansion in $\epsilon$. With this setup, fewer numerical solutions of the integration-by-parts identities are needed in order to reconstruct analytic
results for the amplitude.  This is due to the fact that the expansion of the integrals into pentagon functions, before performing the Laurent
expansion in $\epsilon$ of the final coefficients, complicates the dependence on $\epsilon$ of the result in this intermediate stage.

To make use of the finite field arithmetic we must have a rational parametrisation of the external kinematics.
We parameterise the kinematics using momentum twistors~\cite{Hodges:2009hk,Badger:2013gxa} where,
\begin{equation} \label{eq:s2x}
\begin{aligned}
& s_{12} = x_1 \,, \\
& s_{23} = x_1 x_4 \,, \\
& s_{34} = \frac{x_1}{x_2} (x_4 + x_3 x_4 + x_2 x_3 x_5-x_2 x_3) \,, \\
& s_{45} = x_1 x_5 \,, \\
& s_{15} = x_1 x_3 (x_2-x_4+x_5)\,, \\
& \text{tr}_5 = -\frac{x_1^2}{x_2}\left[ x_2 x_4 (1 + 2 x_3)  - x_4 (1 + x_3)  (x_4 - x_5) + x_2^2 x_3 (-1 + x_5)\right] \, .
\end{aligned}
\end{equation}
We stress that the pseudo-scalar invariant $\text{tr}_5$, and hence the square root of the Gram determinant $\Delta$, is a rational function of the $x_i$ variables. Moreover, since $x_1$ is the only dimensionful variable, we can set it to $1$ and recover the dependence on it after the reconstruction by dimensional analysis. 
Further details on the momentum twistor parameterisation are presented in Appendix~\ref{app:momtwistors}. In the following sections, we will consider all coefficients of the special functions to be rational functions of the variables $x_i$.

\section{Analytic reconstruction over finite fields}
\label{sec:Reconstruction}

In this section, we present three general strategies to optimise the reconstruction over finite fields of the rational coefficients in the finite remainders. At this stage, each component $F(x)$ of the two-loop finite remainder is expressed as
\begin{align} \label{eq:F_r_pf}
F(x) = \sum_{i} r_i(x) \, \text{mon}_i\left(f\right) \,,
\end{align}
where $r_i$ are rational functions of the variables $x$ which parameterise the momentum twistors, and $\text{mon}_{i}(f)$ are linearly independent monomials of the pentagon functions. The entire chain of operations is implemented over finite fields in the framework \finiteflow. We therefore have a numerical algorithm which evaluates the rational coefficients $r_i(x)$ modulo some prime number. The final step consists in reconstructing the analytic expression of the rational coefficients from a sufficient number of numerical evaluations. We employ \finiteflow's multi-variate functional reconstruction algorithms, supplemented with three strategies to reduce the number of required sample points: we determine the linear relations among the rational coefficients and an ansatz, use univariate slices to identify the factors belonging to another ansatz, and perform a univariate partial fraction decomposition on the fly. In the following subsections we discuss thoroughly each of these procedures, and their application to two-loop diphoton finite remainders.

\subsection{Linear relations among the rational coefficients}
\label{sec:LinearRelations}
The representation of the finite remainders in terms of rational coefficients and special function monomials given by Eq.~\eqref{eq:F_r_pf} is in a sense not optimal. The special function monomials in fact do not all appear independently. They are present only in a number of independent combinations that is typically much smaller than the total number of monomials. As a result, the rational coefficients $r_i$ in the finite remainders are not linearly independent. Expressing the finite remainders in terms of a set of linearly independent rational coefficients not only leads to more compact expressions, but may also simplify their reconstruction.

We can determine the linear relations among the rational coefficients $\{r_i(x)\}$ of the special function monomials by solving a linear fit problem,
\begin{align}
\sum_i a_i \, r_i(x)  = 0 \, .
\end{align}
Since the coefficients of the linear relations $a_i$ are rational numbers, they require substantially fewer sample points to be reconstructed with
respect to the rational coefficients themselves. We can then use these relations to express the rational coefficients in terms of a set of linearly
independent ones, which remain to be reconstructed. Choosing the latter to be the simplest ---~i.e.~the ones with the lowest polynomial degrees~--- may reduce the number of sample points required for the reconstruction.
  
This strategy can be further refined by supplying an ansatz for the rational coefficients. We then fit the linear relations among the rational coefficients of the finite remainders and the coefficients of the ansatz, which we denote by $\{e_j(x)\}$,
\begin{align}
\sum_i a_i \, r_i(x) + \sum_j b_j \, e_j(x)  = 0 \, ,
\end{align}
with $a_i, b_j \in \mathbb{Q}$. In the best case scenario, all the rational coefficients $r_i$ can be expressed in terms of the ansatz coefficients $e_j$ and no further reconstruction needs to be performed. Even when the ansatz does not entirely cover the rational coefficients, it may still lower the degrees of the linearly independent coefficients which have to be reconstructed.
The ansatz can be constructed from the tree-level amplitude and the rational coefficients of the one-loop amplitudes up to order $\eps^2$ from the
analysis of the leading singularities~\cite{eden2002analytic,Britto:2004nc,Cachazo:2008vp,ArkaniHamed:2010gh} or from other related amplitudes. In the
diphoton case, we can use the two-loop five-gluon amplitudes. At one loop, the $3g2\gamma$ amplitudes can be expressed in terms of permutations of the five-gluon ones~\cite{Dicus:1987fk,deFlorian:1999tp}. While this is no longer true at two loops, we find there is an important overlap between the rational coefficients of the $3g2\gamma$ amplitudes and those of the five-gluon ones. We use as ansatz in the linear relations the rational coefficients of the leading-colour two-loop five-gluon amplitudes (all two-loop five-parton amplitudes are available analytically at leading colour~\cite{Gehrmann:2015bfy,Dunbar:2016aux,Badger:2018enw,Abreu:2018zmy,Badger:2019djh,Dunbar:2019fcq,Abreu:2021fuk}; we made use of independent results, which are being prepared for publication).

\subsection{Matching factors on univariate slices}
\label{sec:Factors}
The pole structure of the pentagon functions is determined by the letters of the pentagon alphabet~\cite{Chicherin:2017dob}. The pentagon functions
(or their discontinuities) may in fact have logarithmic singularities in the phase-space points where one of the letters vanishes. For this reason, it
is natural to expect that the poles of the rational coefficients should be similarly linked to the pentagon alphabet. Indeed, we observe that the
denominators of the rational coefficients in front of the pentagon functions factorise into a product of letters of the pentagon alphabet. In other words, each rational coefficient $r(x)$ has the form
\begin{align} \label{eq:ansatz_factors}
r(x) = \frac{n(x)}{\prod_{k} {\ell_k}^{e_k}(x)}\,,
\end{align}
where $e_k$ are integers, $n(x)$ is a polynomial in the variables $x$, and $\{\ell_k\}$ is an ansatz of factors from the pentagon alphabet. The exponents $e_k$ in Eq.~\eqref{eq:ansatz_factors} may in general be negative, corresponding to factors in the numerator. We use the following ansatz for the factors,
\begin{equation} \label{eq:coeffansatz}
\begin{aligned}
\left\{\ell_k(x) \right\} = \bigl\{ & \spA 12\,, \spA 13\,, \spA 14\,, \spA 15\,, \spA 23\,,
\spA 24\,, \spA 25\,, \spA 34\,, \spA 35\,, \spA 45\,, \spB 12\,, \spB 13\,, \spB 14\,, \spB 15\,, \\
& \spB 23\,, \spB 24\,, \spB 25\,, \spB 34\,, \spB 35\,, \spB 45\,, 
s_{12} - s_{34}\,, s_{12} - s_{35}\,, s_{12} - s_{45}\,, s_{13} - s_{24}\,, \\
& s_{13} - s_{25}\,, s_{13} - s_{45}\,, s_{14} - s_{23}\,, s_{14} - s_{25}\,, s_{14} - s_{35}\,, s_{15} - s_{23}\,, s_{15} - s_{24}\,, \\
& s_{15} - s_{34}\,, s_{23} - s_{45}\,, s_{24} - s_{35}\,, s_{25} - s_{34}\,, 
 \text{tr}_5 \bigr\}\,.
\end{aligned}
\end{equation}
The exponents $e_k$ in the ansatz~\eqref{eq:ansatz_factors} can be determined by reconstructing $r(x)$ on a univariate slice modulo some prime number~\cite{Abreu:2018zmy}. The univariate slice is defined by parameterising the variables by a single parameter $t$,
\begin{align}
\{ x_i(t) = a_i + b_i t \} \,,
\end{align}
for constant $a_i$ and $b_i$. The latter are chosen randomly in the finite field to avoid artificial simplifications. The dependence on $t$ is chosen to be linear so that the degrees of the numerator and denominator of $r(t):=r\left(x(t)\right)$ correspond to the total degrees of $r$ in $x$. Matching the reconstructed $r(t)$ with the ansatz~\eqref{eq:ansatz_factors} evaluated on the same slice allows to determine the exponents $e_k$ straightforwardly. 
With a univariate reconstruction on just one prime field we can thus infer a lot of information about the analytic form of the rational coefficients: the denominators are entirely fixed, and typically some factors of the numerators are determined as well. What remains to be reconstructed therefore requires fewer sample points.

\subsection{Univariate partial fraction decomposition over finite fields}
\label{sec:Apart}

Partial fraction decomposition is a standard and powerful tool for the simplification of rational functions. The decomposition in partial fractions is however not unique in the multivariate case. Its application to the multivariate rational functions in scattering amplitudes is therefore not straightforward. The necessity to simplify the rational coefficients of two-loop five-particle scattering amplitudes has recently spurred several approaches to handle the multivariate case efficiently~\cite{Abreu:2019odu,Boehm:2020ijp,Heller:2021qkz}, based upon Leinartas' algorithm~\cite{Leinartas:1978,Raichev:2012}. These algorithms rely on algebraic geometry techniques, such as multivariate polynomial division and Gr\"obner bases, and require the arbitrary choice of a monomial ordering. 

Our main goal in this work is actually to simplify the reconstruction of the rational coefficients over finite fields. In other words, we want to reconstruct the rational coefficients on the fly, directly in a form which is decomposed in partial fractions. 
The simplification of the ensuing analytic expressions comes as a welcome by-product. We observe that a univariate partial fraction decomposition is
sufficient for this purpose. The advantage is that it can be straightforwardly implemented over finite fields, avoiding all algebraic geometry
complications. The only arbitrary choice that remains to be done is which variable to partial fraction with respect to. The latter can be chosen by
observing the impact of the partial fraction decomposition with respect to each variable separately on the lower order amplitudes.
With the parameterisation of the kinematics in terms of momentum twistors, Eq.~\eqref{eq:mtparam}, we find it most convenient to partial fraction with respect to $x_4$.

We now discuss our algorithm to reconstruct the univariate partial fraction decomposition of a multivariate rational function $r$ from its numerical evaluations over finite fields. The algorithm requires as input an ansatz for the factors which may appear in the denominator of $r$. Only those factors which depend on the variable with respect to which the partial fraction decomposition is being performed are strictly necessary. Guessing other factors may further simplify the reconstruction. In the application to massless two-loop five-particle scattering amplitudes, the factor ansatz can be inferred from the letters of the pentagon alphabet~\cite{Chicherin:2017dob}. We use the factors in Eq.~\eqref{eq:coeffansatz}.

Let $r$ be a rational function of the variables $x = \{x_i\}_{i=1}^n$. In this work the $x_i$'s are the momentum twistor variables defined by Eq.~\eqref{eq:s2x}, so $n=5$, but we outline the algorithm in general. The goal is to decompose $r$ in partial fractions with respect to one of the variables, say $x_k$. To simplify the notation, we denote the latter by $y = x_k$, and the remaining variables by $\bar{x}=\{x_i\}_{i=1}^n \backslash x_k$. We may not know the analytic expression of $r$, but we must be able to evaluate it numerically modulo some prime number through some algorithm. 
Let $\{ \ell_i(\bar{x},y) \}_{i=1}^m$ be an ansatz for the factors which may appear in the denominator of $r$. Without loss of generality, we assume that the $\ell_i$'s are irreducible polynomials over $\mathbb{Q}$. In other words, we assume that $r$ has the form
\begin{align} \label{eq:Ansatz}
r\left(\bar{x},y\right) = \frac{N\left(\bar{x},y\right)}{\prod_{i=1}^m \ell_i^{e_i}\left(\bar{x},y\right)} \, ,
\end{align}
where $e_i\in \mathbb{Z}$, and $N\left(\bar{x},y\right)$ is a function which depends polynomially on $y$ and rationally on $\bar{x}$. The ansatz $\{ \ell_i\left(\bar{x},y\right) \}_{i=1}^m$ may catch some of the factors in the numerator of $r\left(\bar{x},y\right)$, corresponding to negative values of the exponents $e_i$ in Eq.~\eqref{eq:Ansatz}.  
This lowers the total degrees of $N\left(\bar{x},y\right)$ and eventually simplifies its reconstruction, but is not necessary for the partial fraction decomposition with respect to $y$. Similarly, the ansatz may cover all the factors in the denominator of $r$, so that $N\left(\bar{x},y\right)$ is a polynomial in $\bar{x}$ and $y$. What is necessary for the partial fraction algorithm to work is that the ansatz contains all the factors in the denominator of $r$ which depend on $y$. We denote this subset by
\begin{align}
\Lambda_y = \bigl\{i\in \{1,\ldots,m\} : e_i>0 \land \text{deg}_y\left[\ell_i\left(\bar{x},y\right)\right]>0 \bigr\} \, ,
\end{align}
where $\text{deg}_y\left[h\right]$ is the degree in $y$ of the polynomial $h$.

The first step consists of fixing the exponents $e_i$ in the ansatz~\eqref{eq:Ansatz}. We do this through the procedure discussed in Section~\ref{sec:Factors}.
In the second step we determine the degree in $y$ of the numerator $N\left(\bar{x},y\right)$ in the ansatz~\eqref{eq:Ansatz}. We recall that $N\left(\bar{x},y\right)$ is by construction polynomial in $y$. We compute its degree in $y$ by reconstructing it on another univariate slice, this time where only $y$ varies,
\begin{align} \label{eq:UnisliceY}
\{x_i(t) = a_i \ \forall i\neq k \, ,  \ y(t) = t \} \,,
\end{align}
with $a_i$ chosen randomly in the finite field. Clearly, the degree in $t$ of $N(t):=N\left(\bar{x}=\bar{a},y=t\right)$ gives the degree in $y$ of $N\left(\bar{x},y\right)$. We introduce the short-hand notation
\begin{align}
d_N = \text{deg}_y\left[ N\left(\bar{x},y\right) \right] \,, \quad \qquad d_i = \text{deg}_y\left[ \ell_i\left(\bar{x},y\right) \right] \,, \quad \qquad d_{\Lambda_y} = \sum_{i\in\Lambda_y} e_i d_i 
\end{align}
for the degrees of $N\left(\bar{x},y\right)$ and of the denominator factors $\ell_i\left(\bar{x},y\right)$ in $y$.

Using the information about the factors in the denominator of $r$ and the degree in $y$ of its numerator, we construct the following ansatz for the partial fraction decomposition of $r$ with respect to $y$:
\begin{align} \label{eq:AnsatzPartialFraction}
r\left(\bar{x},y\right) = \sum_{i\in \Lambda_y} \sum_{j=1}^{e_i} \sum_{k=0}^{d_i-1} \frac{U_{ijk}\left(\bar{x}\right) \, y^k}{\ell_i^j\left(\bar{x},y\right)} + R\left(\bar{x}\right) + \sum_{h=1}^{d_N-d_{\Lambda_y}} V_h\left(\bar{x}\right) \, y^h \,,
\end{align}
where $U_{ijk}\left(\bar{x}\right)$, $R\left(\bar{x}\right)$ and $V_h\left(\bar{x}\right)$ are unknown rational functions of $\bar{x}$. 
The right-most term in Eq.~\eqref{eq:AnsatzPartialFraction} is required only if $d_N > d_{\Lambda_y}$, i.e.~only if the numerator of $r$ has a higher degree in $y$ than the denominator. 

The last step of the algorithm consists of reconstructing the analytic dependence on $\bar{x}$ of the unknown coefficients in the
ansatz~\eqref{eq:AnsatzPartialFraction} from the numerical evaluations of $r\left(\bar{x},y\right)$. To solve this linear fit problem, we use the algorithm implemented in the \finiteflow~framework~\cite{Peraro:2019svx}. The solution comes in the form of an algorithm which numerically evaluates $U_{ijk}\left(\bar{x}\right)$, $R\left(\bar{x}\right)$ and $V_h\left(\bar{x}\right)$. 
The rational reconstruction may be simplified by first reconstructing the coefficients on a univariate slice where all the remaining variables $\bar{x}$ vary, and using that to match them with those factors in the ansatz $\{\ell_i\left(\bar{x},y\right)\}_{i=1}^m$ which depend only on $\bar{x}$. This may lower the total degrees of the functions that need to be reconstructed.

In addition to the factors in the original ansatz $\{\ell_i\left(\bar{x},y\right)\}_{i=1}^m$, the coefficients of the partial fraction decomposition~\eqref{eq:AnsatzPartialFraction} may also contain spurious factors. Consider for instance the toy example 
\begin{align} \label{eq:example}
\frac{1}{(y-a)(y-b)} = \frac{1}{(a-b)(y-a)} - \frac{1}{(a-b)(y-b)}  \, ,
\end{align}
where $a$ and $b$ are arbitrary constants such that $a\neq b$. In this example, the inspection of the left-hand side indicates $\{y-a, y-b\}$ as ansatz for the irreducible denominator factors. The partial fraction decomposition however contains a factor of $a-b$ in the denominator, which arises from the residue of the function at the zero of either of the denominator factors. Clearly $a=b$ is a spurious singularity, manifestly absent on the left-hand side and produced by the partial fraction decomposition.
In general, we can determine the potential spurious factors by evaluating the factors in the ansatz $\ell_i\left(\bar{x},y\right)$ which depend on $y$ at their zeros,
\begin{align}
\left\{ \ell_i\left(\bar{x},y^*_k\right)  \right\}_{i\in\Lambda_y, \, k\in\Lambda_{y}^1,\,i\neq k}\,,
\end{align}
where $y^*_k$ is the zero of $\ell_k\left(\bar{x},y\right)$,
\begin{align}
\ell_k\left(\bar{x}, y^*_k\right) = 0\,,
\end{align}
and $\Lambda_{y}^1$ is the subset of factors which depend linearly on $y$,
\begin{align}
\Lambda_{y}^1 = \left\{ i\in \Lambda_y : \text{deg}_y\left[\ell_i\left(\bar{x},y\right)\right]=1\right\}\,.
\end{align}
The restriction to zeros of linear functions of $y$ is due to the facts that the $\ell_i$'s are irreducible polynomials over $\mathbb{Q}$ and that we are factoring over $\mathbb{Q}$. The zeros of higher-degree irreducible polynomials would introduce algebraic and/or complex dependence.

In practice, we observe that determining the spurious factors does not simplify the reconstruction. The greatest part of the denominators of the
coefficients in the partial fraction decomposition~\eqref{eq:AnsatzPartialFraction} is in fact determined by the original ansatz
$\{\ell_i\left(\bar{x},y\right)\}_{i=1}^m$. What remains after they are multiplied away has a total degree which is typically lower than that of the
numerators, which therefore dominates the determination of the number of sample points required for the reconstruction. While it is possible to
determine entirely the denominators of the coefficients in Eq.~\eqref{eq:AnsatzPartialFraction}, it would not reduce the number of required sample points substantially, and for this reason we refrain from doing it.

Having determined as many factors as possible in the coefficients of the partial fraction decomposition, we multiply them away and reconstruct the
remainder using the multivariate rational reconstruction algorithms implemented in \finiteflow. It is important to stress that the algorithm which
evaluates the coefficients of the partial fraction decomposition contains the solution of a linear fit.  For each numerical value of $\bar{x}$,
Eq.~\eqref{eq:AnsatzPartialFraction} is sampled for several numerical values of $y$, roughly as many times as the number of unknowns. This generates a
linear system of equations for the unknowns evaluated at the chosen value of $\bar{x}$. The redundant equations are removed after the learning phase.
The reconstruction on the univariate slices in the intermediate steps of the algorithm, because it requires several evaluations of the original
functions, obviously has a higher computational cost with respect to directly evaluating $r$. On the other hand, the coefficients of the partial
fraction decomposition depend on one fewer variable than the original function $r$, and may have substantially lower degrees.  As a result of all
these aspects, the partial fraction decomposition may be outperformed by a direct reconstruction for simple functions, but becomes more and more
convenient as the complexity of the functions increases.

\subsection{Summary and impact of the reconstruction strategy}
\label{sec:reconstruction_summary}

The techniques discussed in the previous sections are general and can be applied to any rational reconstruction problem, in combination or separately. In order to reconstruct the rational coefficients of the two-loop diphoton finite remainders we apply them consecutively as follows.
\begin{description}
\item[Stage 1.] We fit the linear relations among the rational coefficients with an ansatz, as discussed in Section~\ref{sec:LinearRelations}. We begin with the $(d_s-2)^1$ components and use the coefficients of the two-loop leading-colour five-gluon finite remainders as ansatz. For the $(d_s-2)^0$ components, which are more complicated, we add to the ansatz the $(d_s-2)^1$-coefficients already reconstructed.

\item[Stage 2.] We guess the factors from the ansatz~\eqref{eq:coeffansatz} by reconstructing a univariate slice and multiply them away, as explained in Section~\ref{sec:Factors}. 

\item[Stage 3.] We partial fraction on the fly with respect to $x_4$, applying the algorithm presented in Section~\ref{sec:Apart}. The coefficients to be reconstructed after this stage are those in the ansatz for the partial fraction decomposition~\eqref{eq:AnsatzPartialFraction}, and depend on one fewer variable. 

\item[Stage 4.] We reconstruct another univariate slice and perform an additional factor guessing, as in the second stage.
\end{description}

The drop in the complexity of the rational coefficients after each stage for the most complicated two-loop diphoton finite remainders, which are in the Maximally-Helicity-Violating (MHV) configurations, is illustrated in Table~\ref{tab:degrees_MHV}. As proxy for the complexity of the coefficients we use the maximal numerator/denominator polynomial degrees, which can be evaluated by reconstructing univariate slices as discussed in Section~\ref{sec:Apart}.

\begin{table}
\begin{center}
\begin{tabular}[h!]{ccccccc}
  \hline
  finite remainder & original & stage 1 & stage 2 & stage 3* & stage 4* \\
  \hline
  $F^{(2)}_{1;1}(1^-,2^-,3^+,4^+,5^+)$ & $69/60$ & $28/20$ & $24/0$ & 19/10 & 11/5  \\
  $F^{(2)}_{1;0}(1^-,2^-,3^+,4^+,5^+)$ & $78/69$ & $44/35$ & $43/0$ & $21/10$ & $16/9$ \\
  $F^{(2)}_{1;1}(1^-,2^+,3^+,4^-,5^+)$ & $59/55$ & $30/27$ & $29/0$ & $18/15$ & $17/4$ \\
  $F^{(2)}_{1;0}(1^-,2^+,3^+,4^-,5^+)$ & $89/86$ & $38/36$ & $38/0$ & $20/16$ & $17/3$ \\
  $F^{(2)}_{1;1}(1^+,2^+,3^+,4^-,5^-)$ & $40/42$ & $25/27$ & $25/0$ & $15/18$ & $15/0$ \\
  $F^{(2)}_{1;0}(1^+,2^+,3^+,4^-,5^-)$ & $66/66$ & $32/33$ & $32/0$ & $13/13$ & $12/3$ \\
  \hline
\end{tabular}
\end{center}
\caption{Maximal numerator/denominator polynomial degrees of the rational coefficients of the most complicated finite remainders at each stage of our reconstruction strategy. The column ``original'' refers to the rational coefficients prior to any optimisation. The asterisk * highlights that, after the partial fraction decomposition in stage 3, the coefficients to be reconstructed depend on one fewer variable.
}
\label{tab:degrees_MHV}
\end{table}

Interestingly, we observe that the coefficients of the subleading-colour $3g2\gamma$ two-loop finite remainders $F^{(2)}_2$ can be expressed in terms of those of the leading-colour two-loop five-gluon finite remainders. The coefficients of the leading-colour $3g2\gamma$ two-loop remainders $F^{(2)}_1$ instead are not entirely fixed by the five-gluon ones, but using the latter as ansatz in the linear relations reduces significantly the maximal polynomial degrees of the coefficients which remain to be reconstructed.

As can be appreciated in Table~\ref{tab:degrees_MHV}, our strategy leads to a substantial drop in the polynomial degrees. Furthermore, the coefficients to be reconstructed after the partial fraction decomposition (stage 3) depend on one fewer variable. This makes the decrease in the number of sample points required for the reconstruction even more pronounced.
The price to pay for this is that performing the partial fraction decomposition increases the evaluation time per point, as discussed at the end of Section~\ref{sec:Apart}.
With our setup we observe that, for the most complicated finite remainders, the evaluation times grows roughly by one order of magnitude, while the
number of sample points required for the reconstruction decreases by two orders of magnitude. This leads to an overall gain of roughly one order of magnitude in the
reconstruction time. We stress that the evaluation time relevant here is that of the algorithm which evaluates the rational coefficients over finite
fields, not the final evaluation time of the finite remainders. Once the reconstruction is completed, in fact, the rational coefficients are evaluated
from their analytic expressions. For the evaluation time of the finite remainders, we refer to Section \ref{sec:performance}.

Our approach therefore leads to an important simplification in the reconstruction of the rational coefficients. Moreover, the ensuing analytic
expressions are dramatically more compact. This makes them suitable for compilation in a \cpp~library, an essential step for future phenomenological applications.

\section{Compact analytic expressions for the all-plus configuration}
\label{sec:AllPlus}

Prior to discussing the numerical implementation of all two-loop helicity amplitudes, we would like to comment on the all-plus amplitude, which displays a particularly simple analytic form. We find that the structures appearing are closely related to those appearing in the five-gluon all-plus amplitudes at one~\cite{Bern:1993sx,Mahlon:1993si,Bern:1993qk,Henn:2019mvc} and two loops~\cite{Gehrmann:2015bfy,Dunbar:2016aux,Badger:2019djh,Dunbar:2019fcq}. We present the finite remainders in the expansion around $d_s=2$. 

The all-plus amplitude is finite and rational at one loop. The finite remainder can be written as
\begin{align}
  F_{1;0}^{(1)}(1^+,2^+,3^+,4^+_\gamma,5^+_\gamma) = - 2 \frac{\spB45^2}{\spA12\spA23\spA31} \, .
  \label{eq:F1allplus}
\end{align}
Remarkably, this amplitude is invariant under conformal transformations, and the expression given here exhibits this property in a manifest way~\cite{Henn:2019mvc}. If all masses are neglected, the SM Lagrangian is conformally invariant. This symmetry is obscured at loop level by the appearance of scales associated with the divergences and it is therefore rather surprising to observe it in a one-loop amplitude. One might na\"ively suppose that this is a consequence of the finiteness of the all-plus one-loop amplitudes. Yet, the single-minus one-loop amplitudes are equally finite, but they are not conformally invariant. This phenomenon still calls for an explanation. These properties are discussed in detail in Ref.~\cite{Henn:2019mvc}, where the authors prove that the $n$-gluon amplitudes in QCD are conformally invariant at one loop. Since the diphoton amplitudes can be expressed as permutations of pure gluon scattering~\cite{Dicus:1987fk,deFlorian:1999tp} and the conformal generators commute with permutations, all considerations regarding conformal symmetry trivially extend to the diphoton case.

At two-loop order, the $d_s=2$ contribution is the only one involving transcendental functions. Its expression is remarkably simple, 
\begin{align}
  F_{1;0}^{(2)}(1^+,2^+,3^+,4^+_\gamma,5^+_\gamma) = \frac{\spB45^2}{\spA12\spA23\spA31} \sum_{{\rm cyclic}(123)} F_{\rm box}(s_{12},s_{23};s_{45}) \, ,
  \label{eq:F2allplusA}
\end{align}
where the sum runs over the cyclic permutations of $(1,2,3)$, and
\begin{align}
F_{\rm box}(s_{12},s_{23};s_{45}) = {\rm Li}_2\left(1-\frac{s_{12}}{s_{45}}\right) + {\rm Li}_2\left(1-\frac{s_{23}}{s_{45}}\right) + \log^2\left(\frac{s_{12}}{s_{23}}\right) + \frac{\pi^2}{6} 
\end{align}
is the finite part of the one-loop box with an off-shell leg. The analytic continuation of the box functions to any scattering region can be easily achieved by adding a small positive imaginary part to each two-particle Mandelstam invariant, $s_{ij} \to s_{ij} + i0^+$. The other partial amplitudes at two loops are rational,
\begin{align}
  F_{1;1}^{(2)}(1^+,2^+,3^+,4^+_\gamma,5^+_\gamma) &=
      - \frac{\spB45^2}{\spA12\spA23\spA31}
      - \frac{1}{2} F_{3;0}^{(2)}(1^+,2^+,3^+,4^+_\gamma,5^+_\gamma) \,, \nonumber\\
  F_{2;0}^{(2)}(1^+,2^+,3^+,4^+_\gamma,5^+_\gamma) &= 0 \,, \nonumber\\
  F_{2;1}^{(2)}(1^+,2^+,3^+,4^+_\gamma,5^+_\gamma) &=
      - 3 \frac{\spB45^2}{\spA12\spA23\spA31}
      - \frac{1}{2} \frac{\tr_5\left(p_1,p_2,p_3,p_4-p_5\right)\spA45}{\spA14\spA15\spA24\spA25\spA34\spA35} \,,  \nonumber\\
  F_{3;0}^{(2)}(1^+,2^+,3^+,4^+_\gamma,5^+_\gamma) &=
   \frac{1}{3} \tr_5\left(p_1,p_2,p_3,p_4-p_5\right) \sum_{{\rm cyclic}(123)} \frac{1}{\spA23^2\spA14\spA15 \spA45}\,,
  \label{eq:F2allplusB}
\end{align}
where $\tr_5(p_i,p_j,p_k,p_l) = \tr(\gamma_5 \slashed{p}_i \slashed{p}_j \slashed{p}_k \slashed{p}_l )$. The peculiar simplicity of this amplitude at
two loops follows from the fact that it vanishes at tree level and it is rational in four dimensions at one loop. The one-loop amplitude can in fact
be used as an effective on-shell vertex in four-dimensional unitarity~\cite{Bern:1994zx,Bern:1994cg,Dunbar:2016aux}.
In this way, the cuts of the two-loop amplitude become one-loop cuts with an insertion of the effective vertex. The one- and two-loop all-plus finite remainders are thus treated as tree-level and one-loop objects, respectively. As a result, the special functions appearing in the finite remainder at two loops can have at most transcendental weight two (up to order $\eps^0$). Moreover, the rational coefficients of the transcendental functions can be shown through four-dimensional unitarity to be given by (permutations of) the one-loop all-plus finite remainder. They thus inherit the symmetry under conformal transformations from the one-loop amplitude. These beautiful properties are manifest in our explicit expressions~\eqref{eq:F2allplusA} and~\eqref{eq:F1allplus}. Complementing four-dimensional unitarity with recursion relations for the rational terms allows to compute the two-loop all-plus finite remainders in the purely gluonic case avoiding altogether the computation of the two-loop integrals~\cite{Dunbar:2016aux,Dunbar:2019fcq}. Some results are available even for amplitudes involving more than five plus-helicity gluons~\cite{Dunbar:2016cxp,Dunbar:2016gjb,Badger:2016ozq,Dunbar:2017nfy,Dunbar:2020wdh,Dalgleish:2020mof}.

Amplitudes with a single minus helicity share some of the simplicity of the all-plus case. They also vanish at tree level, and are finite and rational at one loop. As a result, they also have maximum transcendental weight two at two loops. Differently from the all-plus amplitudes, however, they do not have the structure that $F_{1;0}^{(2)}$ has uniform transcendental weight two with all other contributions being
rational. For the amplitudes with two negative helicities, instead, the finite remainders have maximum weight two and four at one and two loops, respectively.

\section{Implementation and performance}
\label{sec:performance}

The finite remainders are coded up into the \njet~\cpp~library, which is linked to the \pentagonfunctions~library~\cite{Chicherin:2020oor} for the evaluation of the special functions.
The six independent helicity amplitudes (shown in Table~\ref{tab:finremeval}) are permuted analytically onto the global basis of pentagon functions defined in the $12\to345$ scattering region to provide a complete list of 16 ``mostly-plus'' helicity amplitudes required for the sum.
This task is performed using the permuted coefficients from the six fully reconstructed amplitudes as an ansatz into the linear relations so additional reconstruction time is avoided (see Section~\ref{sec:LinearRelations}).
Having identified a global basis of pentagon functions for the complete colour and helicity sum, we formulate the partial amplitudes as
\begin{equation}
  F^{h} = c^h_{i} M^h_{ij} f^h_j \,,
  \label{eq:partialrep}
\end{equation}
where $h$ is the helicity configuration, $f^h_j$ is a list of integers corresponding to the global list of pentagon function monomials,
$M^h_{ij}$ are sparse matrices of rational numbers,
and $c^h_{i}$ are the independent rational coefficients written in terms of independent polynomials in the momentum twistor variables $x_i$.
The pentagon function monomials are split into parity-odd and -even components, which allows the remaining 16 ``mostly-minus'' helicities to be computed by simply flipping the parity of the special functions and applying complex conjugation to the coefficients.
The colour- and helicity-summed matrix element is constructed numerically from these ingredients.
The sparse matrix multiplication is implemented using the \eigen~library~\cite{eigenweb}.
Evaluation with 128-bit and 256-bit floating-point numbers (\fpp~and \fppp) is provided via the \qd~library~\cite{qdweb}.
The code is available through \href{https://bitbucket.org/njet/njet}{\texttt{https://bitbucket.org/njet/njet}}, where we provide additional installation instructions and example programs demonstrating the usage.

The \texttt{C++} code returns the values of the one- and two-loop hard functions, $\mathcal{H}^{(1)}$ and $\mathcal{H}^{(2)}$, obtained by squaring Eq.~\eqref{eq:colourdecomp}, substituting the decomposition in $N_c$ and
$n_f$ from Eq.~\eqref{eq:Nc_exp}, subtracting the IR and UV poles, and finally summing over colour and helicity,
\begin{equation}
\begin{aligned}  
  \mathcal{H} &= \frac{\alpha^2 \alpha_s^3}{(4\pi)^5} \left(
  \mathcal{H}^{(1)} + \frac{\alpha_s}{4\pi} \, \mathcal{H}^{(2)}
  \right) + \mathcal{O}(\alpha_s^5)\,,\\
  \mathcal{H}^{(2)} &= 
  N_c \, \mathcal{H}^{(2)}_1
  + \frac{1}{N_c} \, \mathcal{H}^{(2)}_2
  + n_f \, \mathcal{H}^{(2)}_3\,.
  \label{eq:hardfunction}
\end{aligned}  
\end{equation}
The sum over colours for each helicity can also be returned if required. We find the evaluation time is dominated by the special functions,
particularly when higher precision is required. In order to ensure fast and
stable numerical evaluation, we adopt the following evaluation strategy.
\begin{enumerate}
  \item The user-provided phase-space point is checked for the precision of the on-shell constraints. Points are adjusted in case the precision is not
    acceptable for the requested number of digits: 64-bit floating-point numbers (\fp) $\sim \num{15}$; \fpp~$\sim \num{31}$; \fppp~$\sim \num{62}$.
  \item The colour- and helicity-summed amplitude is computed using \fp~precision at two points which differ only by overall dimension scaling
    factor. After accounting for the overall dimension of the squared amplitude, the two evaluations should only differ due to
    rounding errors at intermediate stages in the evaluation of the coefficients. This accuracy scaling test has been used extensively
    at one loop. We refer to this accuracy as \fp/\fp~since both coefficients and special functions use \fp~precision.
  \item If the estimated number of correct digits from the scaling tests falls below a user-defined threshold, the coefficients only are recomputed
    using \fpp~precision after the original point is corrected to \fpp~precision (as in step 1). We refer to this as \fpp/\fp~precision.
  \item The scaling test is performed again and if it fails the special functions are re-evaluated in \fpp~precision. This is \fpp/\fpp~precision.
  \item These steps can be repeated to obtain up to \fppp/\fppp~precision. In practice these steps are rather expensive and unnecessary for standard phenomenological applications, so they are omitted from our strategy.
\end{enumerate}

While the dimension scaling test has been used successfully at one loop, we need to be more careful in our applications when linking the \pentagonfunctions~library, which also makes use of the dimension re-scaling internally.
To validate the reliability of the scaling test as an estimate of the error of the result, we evaluate both with a direct \fpp/\fpp~computation and
via a scaling test with an error cutoff of four digits at \fp/\fp~for a set of \num{60000} points. 
To ensure a realistic validation, we use ``physical'' points with a phase-space sampling density determined by the one-loop process, obtained from \nnlojet.
We compare the estimated error provided by the \fp/\fp~scaling test to the relative difference between the \fp/\fp~and \fpp/\fpp~evaluations, which is taken as the true error.
In the following, percentages are always with respect to the entire set of points.

The scaling test returns a negative for \SI{2.8}{\percent} of the points.
According to true error, an additional \SI{0.2}{\percent} of the points should be failed and are missed by the scaling test (false positive).
Of these points, almost all have true error of four digits, the remaining \SI{0.008}{\percent} with three digits, so the effect on stability is small.
The scaling test also fails some points unnecessarily (false negative), this subset comprising \SI{0.7}{\percent} of all points, which incurs a small performance penalty in the evaluation strategy.
The effects of the false estimates are considered to be allowably small.

We note that the dimension scaling test is statistical and therefore one will always find anomalies in a sufficiently large sample.
Care should be taken when integrating over extreme regions of phase space.

\begin{figure}
\centering
\includegraphics[width=10cm]{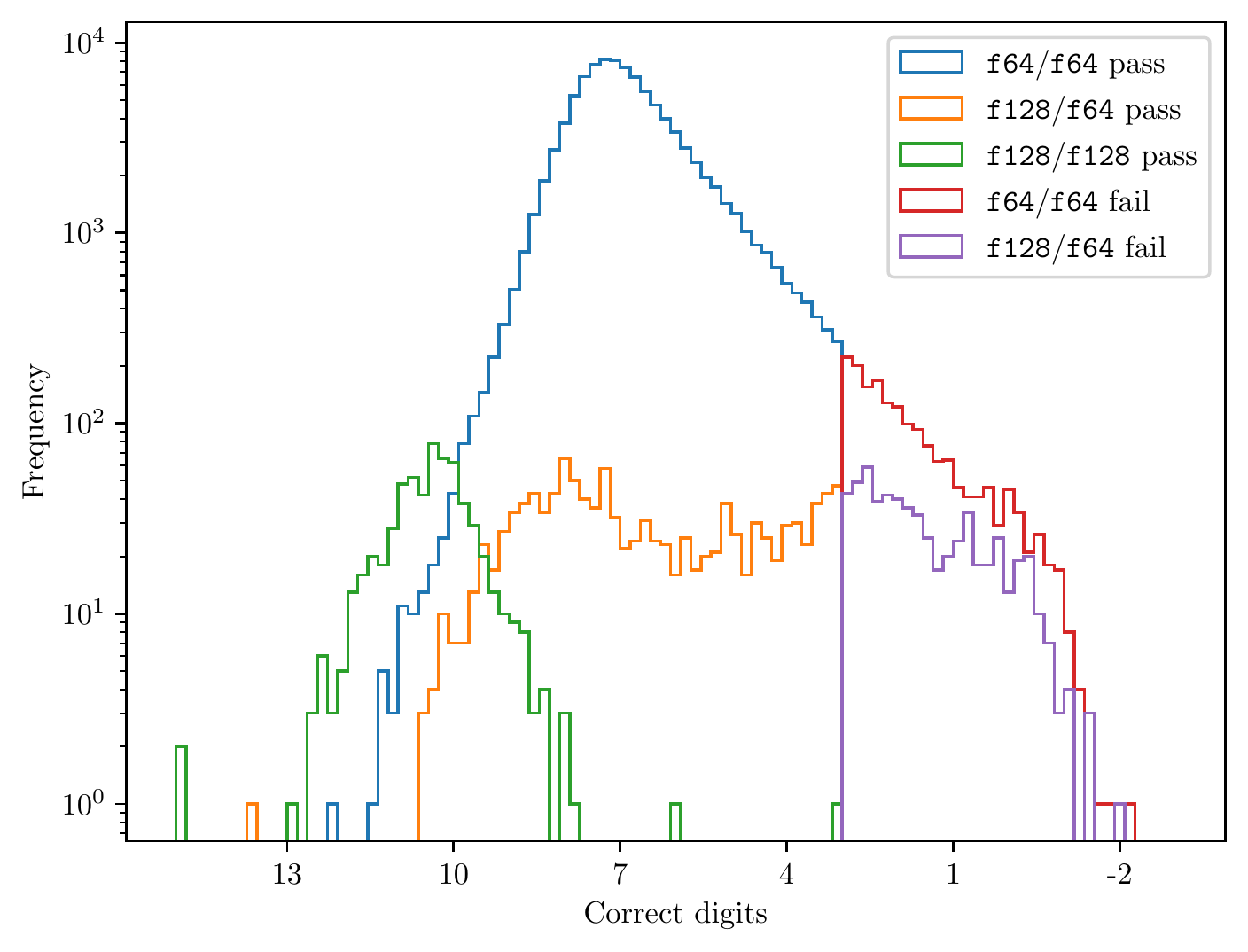}
\caption{
    Histogram of the error estimate on the two-loop evaluations as given by the scaling test.
    We use the evaluation strategy with a target accuracy of three digits and show errors for all precision levels.
    We see \SI{1.8}{\percent} of points failing \fp/\fp~evaluation, with \SI{1.2}{\percent} passing at \fpp/\fp~and \SI{0.6}{\percent} passing at \fpp/\fpp.
    The evaluation strategy achieves target accuracy for all of the \num{100000} physical phase-space points tested.
}
\label{fig:stability}
\end{figure}

To assess the stability of our implementation (Figure~\ref{fig:stability}) and measure timings, we evaluate the amplitude squared over \num{100000}
points of the physical phase space.
We find a single \fp/\fp~call has a mean time of \SI{9}{\second}, with \SI{99}{\percent} of that time spent evaluating the pentagon functions.
Using the full evaluation strategy with a target minimum accuracy of three digits, we obtain a mean timing per phase-space point of \SI{26}{\second}.

We present a benchmark evaluation at a point taken from the physical phase space.
We choose a generic configuration where the momentum invariants ($\si{GeV}^2$) and $\mathrm{tr}_5$ ($\si{GeV}^4$) take the values, quoted to four
significant figures,
\sisetup{
    round-mode = figures,
    scientific-notation = fixed,
}
\begin{equation}
\begin{aligned}  
s_{12} &= \num{1.4116251163350877377236204264908242298667471893246336180766098423e+04} \,, &
s_{23} &= \num{-1.4046834737972321415660577595055706351599646008939620136601159902e+03} \,, &
s_{34} &= \num{7.6669799448946075091609646313550363122719368647199176601634259772e+03} \,, \\
s_{45} &= \num{5.4932450565561162032281304796011000981320172481749213347830057155e+03} \,, &
s_{15} &= \num{-4.4044289245917486109857445869899023397535714857496713286053642996e+03} \,, &
\mathrm{tr}_5 &= \num{-1.7599755750687916479576367229182851347753538557888213932493870876e+07}\mathrm{i}\,.
\label{eq:benchmarkpoint}
\end{aligned}
\end{equation}

\sisetup{
    round-precision = 1,
    scientific-notation = false,
}

\noindent High precision \fpp/\fpp~evaluations are given in the ancillary files.
The values for the finite remainders and the two-loop hard function, normalised by the leading order, are shown in Tables \ref{tab:finremeval} and \ref{tab:hardfuneval} respectively.
The subleading-colour corrections are \num{645} times smaller than the leading colour at this point, while the closed fermion loop corrections
are \num{133} times smaller.
These ratios do change as we sample different points.
Averaging over \num{100} physical points, the ratio
$\lvert N_c\mathcal{H}^{(2)}_1/\mathcal{H}^{(1)}\rvert:\lvert\frac{1}{N_c}\mathcal{H}^{(2)}_2/\mathcal{H}^{(1)}\rvert:\lvert n_f\mathcal{H}^{(2)}_3/\mathcal{H}^{(1)}\rvert$
is \num{2061}:\num{1}:\num{14}.

While the evaluation is considerably more difficult than the massless planar five-gluon scattering owing to the more complicated set of pentagon
functions, our tests show the amplitudes are clearly ready for phenomenological applications.

\sisetup{
    round-precision = 4,
    scientific-notation = fixed,
}

\begin{table}
  \centering
  \begin{tabular}[h]{cccc}
    \hline
    helicity & $N_c F^{(2)}_1/F^{(1)}$ & $\frac{1}{N_c} F^{(2)}_2/F^{(1)}$ & $n_f F^{(2)}_3/F^{(1)}$ \\
    \hline
    $+++++$ & 
        $\num{-2.77582993046583427605142515235285e+01} - \num{1.01745076290502519208036374477010e+01}$i & 
        $\num{-1.67327484248907239815459032713580e+00} - \num{2.39649153322601693283371805022356e-01}$i & 
        $\num{-5.22837426806977838213461692255251e+00} - \num{4.03428197735324580204112002553437e+00}$i \\
    $-++++$ & 
        $\num{-2.57580532623280503092540771294086e+01} + \num{2.78347349175690890510559905150441e+01}$i & 
        $\num{3.57081631633550999314235864904020e-01} - \num{3.21338918788382341155048018442021e-01}$i & 
        $\num{3.36268118497271552633054996889330e-01} - \num{4.42401378002389410311387525774025e+00}$i \\
    $+++-+$ &
        $\num{-2.41566460358473862557597565808062e+01} + \num{1.45880275249122297539666099129978e+01}$i &
        $\num{3.69756140499054359520282727639744e-01} - \num{5.53932445125758408476522017564549e-01}$i &
        $\num{-4.95123441825520347867841272615952e+00} + \num{6.67167689442155278362227939920601e-01}$i \\
    $--+++$ &
        $\num{-2.02342531062579671222860854129219e+01} + \num{8.20384686558306263565400583003510e-01}$i &
        $\num{-4.05487410679291377838146170351151e-01} - \num{3.54867788156361451381190048399781e-01}$i &
        $\num{5.35518188040192554893963855293205e-02} + \num{2.47830230193615584923072985322998e-04}$i \\
    $-++-+$ &
        $\num{-2.85844161095369508004626750797306e+01} + \num{3.29007964718656854030254373989609e+01}$i &
        $\num{3.91732905857814948144244362925973e-01} - \num{5.48869824722050041995728687006486e-04}$i &
        $\num{3.02156400028686127902143151309653e+00} + \num{1.47549772885303042900035952799006e+00}$i \\
    $+++--$ &
        $\num{-2.09417707770365479210244429535450e+01} - \num{1.53427303145404829662973631186851e+01}$i &
        $\num{-3.07968164685809980488634602515262e-01} - \num{4.55762937816688321755049623598366e-01}$i &
        $\num{-4.87984856920121200597547643690781e+00} - \num{5.86222628879890415882513441668661e-03}$i \\
    \hline
  \end{tabular}
  \caption{Numerical values of the partial amplitudes for the six independent helicities at the benchmark point in Eq.~\eqref{eq:benchmarkpoint}.
      Values are quoted with $N_c=3$ and $n_f=5$, to four significant figures.}
  \label{tab:finremeval}
\end{table}

\begin{table}
  \centering
  \begin{tabular}[h]{ccc}
    \hline
    $N_c\mathcal{H}^{(2)}_1/\mathcal{H}^{(1)}$ &
    $\frac{1}{N_c}\mathcal{H}^{(2)}_2/\mathcal{H}^{(1)}$ &
    $n_f\mathcal{H}^{(2)}_3/\mathcal{H}^{(1)}$ \\
    \hline
    $\num{52.74593955908158767014772455004}$ &
    $\num{0.08176280908057119497025187026506}$ &
    $\num{0.39564191164271056748420506863181}$ \\
    \hline
  \end{tabular}
  \caption{
      Numerical values for the components of the two-loop hard function normalised to the one-loop hard function defined in
      Eq.~\eqref{eq:hardfunction} at the benchmark point of Eq.~\eqref{eq:benchmarkpoint}.
      Values are quoted with $N_c=3$ and $n_f=5$, to four significant figures.}
  \label{tab:hardfuneval}
\end{table}

\sisetup{
    round-mode = off,
    scientific-notation = false,
}

\section{Conclusions}

In this paper we have presented a complete, full colour, five-point amplitude at two loops in QCD. All helicity configurations have been implemented into the \njet~\cpp~library, which provides an efficient and stable evaluation over the physical scattering region.
Though the algebraic complexity of the amplitude is considerable, the direct analytic reconstruction of the finite remainders was possible by making
use of linear relation amongst the coefficients and partial fractioning in one variable, which could be done without any analytic knowledge of the
intermediate steps in the reduction. We expect these techniques will have applications to other important high-multiplicity two-loop calculations with more external scales such as five-particle scattering with an off-shell leg, for which there has also been recent
progress~\cite{Hartanto:2019uvl,Papadopoulos:2015jft,Papadopoulos:2019iam,Abreu:2020jxa,Canko:2020ylt,Syrrakos:2020kba,Badger:2021nhg}. We have found a form that is suitable for phenomenological applications and look forward to new precision predictions for diphoton production at hadron colliders including the dominant N${}^3$LO corrections we have computed here.

\acknowledgments
We are grateful to Vasily Sotnikov for useful discussions. This project received funding from the European Union's Horizon 2020 research and innovation programmes
\textit{New level of theoretical precision for LHC Run 2 and beyond} (grant agreement No 683211),
\textit{High precision multi-jet dynamics at the LHC} (grant agreement No 772009), and
\textit{Novel structures in scattering amplitudes} (grant agreement No 725110), 
and from the Swiss National Science Foundation (SNF) under contract 200020-175595.
HBH was partially supported by STFC consolidated HEP theory grant ST/T000694/1.
DC is supported by the French National Research Agency in the framework of the ``Investissements d’avenir'' program (ANR-15-IDEX-02).
RM was supported by STFC ST/S505365/1 and ST/P001246/1.
SZ gratefully acknowledges the computing resources provided by the Max Planck Institute for Physics and by the Max Planck Computing \& Data Facility.

\appendix

\section{Momentum twistor parametrisation \label{app:momtwistors}}

Following~\cite{Hodges:2009hk,Badger:2013gxa,Badger:2017jhb}, the construction begins with
\begin{align}
Z_i = \begin{pmatrix} \lambda_i \\ \mu_i \end{pmatrix} \, ,
\end{align}
where $\lambda_i$ is the negative-helicity spinor, and $\mu_i$ is related to the positive-helicity spinor $\tilde{\lambda}_i$ via
\begin{align}
\tilde{\lambda}_i = \frac{\langle i, i+1 \rangle \, \mu_{i-1} + \langle i+1, i-1 \rangle \, \mu_{i} + \langle i-1, i \rangle \, \mu_{i+1}  }{\langle i, i+1 \rangle \langle i-1, i \rangle} \, ,
\end{align}
with the indices defined modulo $5$. Using the Poincar\'e and $U(1)$ symmetries it is possible to fix all but $5$ of the entries
of the momentum twistor matrix $Z = (Z_i)_{i=1,\ldots,5}$. Explicitly we choose the form,
\begin{align}
Z = \begin{pmatrix} \lambda_i \\ \mu_i \end{pmatrix}_{i=1,\ldots,5} \, = \begin{pmatrix}
    1 & 0 & \frac{1}{x_1} & \frac{1+x_2}{x_1 x_2} & \frac{1+x_3 (1+x_2)}{x_1 x_2 x_3} \\
    0 & 1 & 1 & 1 & 1 \\
    0 & 0 & 0 & \frac{x_4}{x_2} & 1 \\
    0 & 0 & 1 & 1 & \frac{x_4-x_5}{x_4}
    \label{eq:mtparam}
\end{pmatrix} \,,
\end{align}
The parameterisation used in this work has some benefits: the only dimensionful quantity is
$x_1$ and all holomorphic quantities are described using only $x_1,x_2,x_3$. For real kinematics only $x_2$ and $x_3$ are complex. Notice that the conversion between the momentum twistor coordinates and spinor-helicity expressions is only invertible for phase-free quantities.
For this purpose we may use the following relations,
\begin{equation} \label{eq:x2s}
\begin{aligned}
& x_1 = s_{12} \,, \\
& x_2 =  - \frac{\text{tr}_+(p_1,p_2,p_3,p_4)}{s_{12} s_{34}} \,, \\
& x_3 =  - \frac{\text{tr}_+(p_1,p_3,p_4,p_5)}{s_{13} s_{45}} \,, \\
& x_4 = \frac{s_{23}}{s_{12}} \,, \\
& x_5 = \frac{s_{45}}{s_{12}} \,,
\end{aligned}
\end{equation}
with $\text{tr}_+(p_i,p_j,p_k,p_l) = \text{tr}[(1+\gamma_5)\slashed{p}_i \slashed{p}_j \slashed{p}_k \slashed{p}_l ]/2 = \spB ij \spA jk \spB kl \spA li$. 

In our work we express the helicity amplitudes in terms of the momentum twistors variables $x_i$. The phase information can be restored by multiplying and dividing by a suitable phase factor,
\begin{align} \label{eq:phase_normalisation}
\mathcal{A} = \Phi(\lambda_i, \tilde{\lambda}_i) \, \left( \frac{\mathcal{A}(x_i)}{\Phi (x_i)} \right)\,,
\end{align}
where $\mathcal{A}$ is an helicity amplitude --~or in general some object with a non-trivial phase~-- and $\Phi$ is an arbitrary factor with the same helicity weights as $\mathcal{A}$. The quantities in the parentheses in Eq.~\eqref{eq:phase_normalisation} are both written in terms of momentum twistors. Their ratio is phase-free and can thus be expressed in terms of the scalar and pseudo-scalar invariants $s_{ij}$'s and $\text{tr}_5$, e.g. through Eqs.~\eqref{eq:x2s}. The factor outside the parenthesis is instead written in terms of the spinor helicity variables and carries all the phase information of $\mathcal{A}$.

\bibliography{3g2a.bib}
\bibliographystyle{JHEP}

\end{document}